\documentclass[a4paper]{jpconf}
\usepackage{amssymb}
\usepackage{graphicx}
\newcommand{\Order}{\mathcal{O}}
\begin{document}
\title{The Beauty of Spin}

\author{Ulf-G. Mei\ss ner}

\address{Universit\"at Bonn, HISKP and Bethe Center for Theoretical Physics,
  D-53115 Bonn, Germany\\ 
  Forschungszentrum J\"ulich, IKP-3, IAS-4 and JCHP, D-52425 J\"ulich, Germany}

\ead{meissner@hiskp.uni-bonn.de}

\begin{abstract}
I review recent developments in theoretical spin physics. Topics include
pion production in nucleon-nucleon collisions, the implications of heavy quark
spin symmetry for heavy hadron molecules, the nucleon electric dipole form
factors and  ab initio calculations of the width of hadron resonances. A few
spin physics high-lights from experiments at the COSY accelerator are also discussed.

\end{abstract}

\section{Introduction and disclaimer}

Spin is an intrinsic quantity of fundamental and composite particles.
It is particularly fascinating since it is a quantum phenomenon -
spin simply does not exist in classical physics. Furthermore,
relativity combined with quantum mechanics leads to the spin-statistics
theorem, which is yet another intriguing aspect of spin.
Spin is also one of the finest experimental tools as it entails polarization.
More precisely, experiments with polarized beams and/or targets allow to
test the inner workings of the Standard Model, as will be discussed in
a few selected examples below. Spin also entails a magnetic moment, 
and the comparison of its calculation and measurements has led to some
of the most precise tests in physics, like e.g. for QED (see the recent
paper \cite{Aoyama:2010pk} and references therein). Here, I can only address some topics 
related to spin and leave the discussion on the spin of the proton, the measurements
and interpretation of single-spin asymmetries, the role of spin-dependent nuclear
forces, or double polarization experiments at ELSA (Bonn), just to name a few,  
to other speakers at this symposium. In particular,
an experimental overview on recent achievements in spin physics is given by
Milner~\cite{RM}.

\section{Theoretical spin physics at J\"ulich: a short overview}

First, I would like to review some recent work on theoretical spin
physics performed at J\"ulich. This is, of course, a very subjective
choice but is intended to underline the importance of the spin degrees of
freedom in hadron and nuclear physics.

\begin{figure}[t]
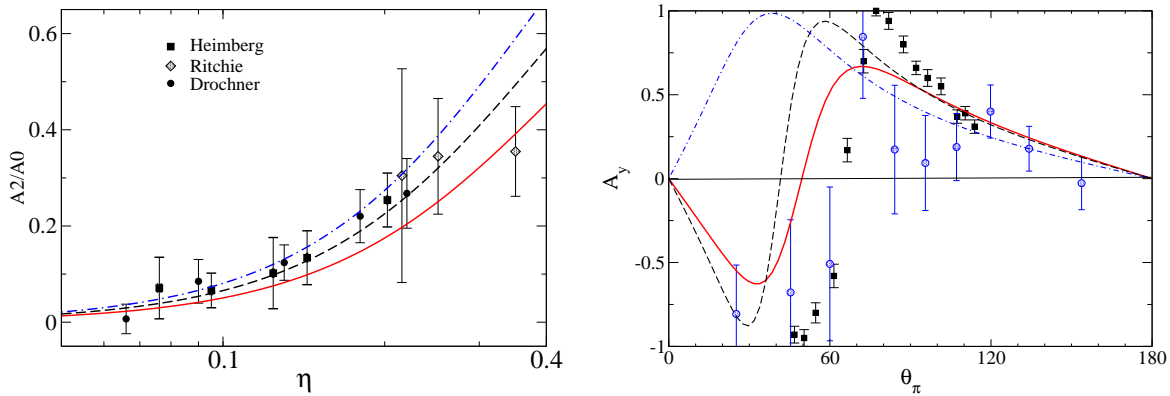

\begin{minipage}{36pc}
\vspace{4mm}
\includegraphics[width=0.48\textwidth]{A2_A0_paper.eps}\hspace{1pc}
\includegraphics[width=0.49\textwidth]{Ay_paper_with_pf_final.eps}
\vspace{-2mm}
\caption{\label{fig:NNpi}Left: Results for the ratio of the Legendre
coefficients in the threshold region for $pp\to d\pi^+$ as a function
of the strength of the LEC $D$. $D=3$: solid, $D=0$: dashed, $D=-3$:
dot-dashed. Here, $\eta$ is the pion momentum in  units of the pion 
mass. Data: Ref.~\cite{A0A2data}. Right: Analyzing power $A_y$ for 
$pn\to pp(^1S_0)\pi^-$. Data: Ref.~\cite{Aydata}.}
\end{minipage}\hspace{2pc}%
\end{figure}

Consider first the light quark sector of QCD, i.e. the two-flavor world
of up and down quarks. Here, chiral perturbation
theory (or related effective field theories) is the appropriate
theoretical tool. Many processes have been analyzed in this
framework, see e.g. Ref.~\cite{Bernard:2007zu}. The extension 
to systems with two or more nucleons requires some additional
non-perturbative resummation, for a review see~\cite{Epelbaum:2008ga}. 
Recently, a combined analysis of the
reactions  $pn\to pp\pi^-$, $pp\to pn\pi^+$ and $pp\to d\pi^+$ at 
next-to-next-to-leading order (NNLO) in the chiral expansion became
available~\cite{Baru:2009fm}. In particular, the existing data were
analyzed as a function of the three-nucleon low-energy constant (LEC) $D$
that parameterizes the leading $(\bar NN)^2\pi$ contact operator. It
features prominently in a variety of low-energy processes like
$Nd \to Nd$, $NN\to NN\pi$, $NN\to d l \nu_l$, $\gamma d \to NN\pi$ 
or $d\pi\to NN\gamma$. A precise determination of this LEC would therefore lead to stringent
tests of the chiral QCD dynamics. In pion production reactions, one
has to account for the scale $\sqrt{M_\pi m}\simeq 340\,$MeV, with 
$M_\pi\, (m)$ the pion (nucleon) mass, in setting up the power counting,
see \cite{Hanhart:2003pg} for a review.
As can be seen from Fig.~\ref{fig:NNpi}, the existing data are not precise enough to
pin down $D$ accurately, but seem to favor a positive value. Note that,
not unexpected, the polarization observable $A_y$ (right panel) is more
sensitive to $D$ than the threshold differential cross section, evaluated
as $d\sigma/d\Omega = A_0 +A_2 P_2(\cos\theta_\pi)$ (left panel) (for more
details, see \cite{VB}). Therefore,
at COSY an experiment has been proposed to determine $D$ accurately
from the  measurement of $\vec p\, \vec n \to \{pp\}_s \pi^-$~\cite{dymov}.

\begin{figure}[tb]
\begin{minipage}{36pc}
\includegraphics[width=0.46\textwidth]{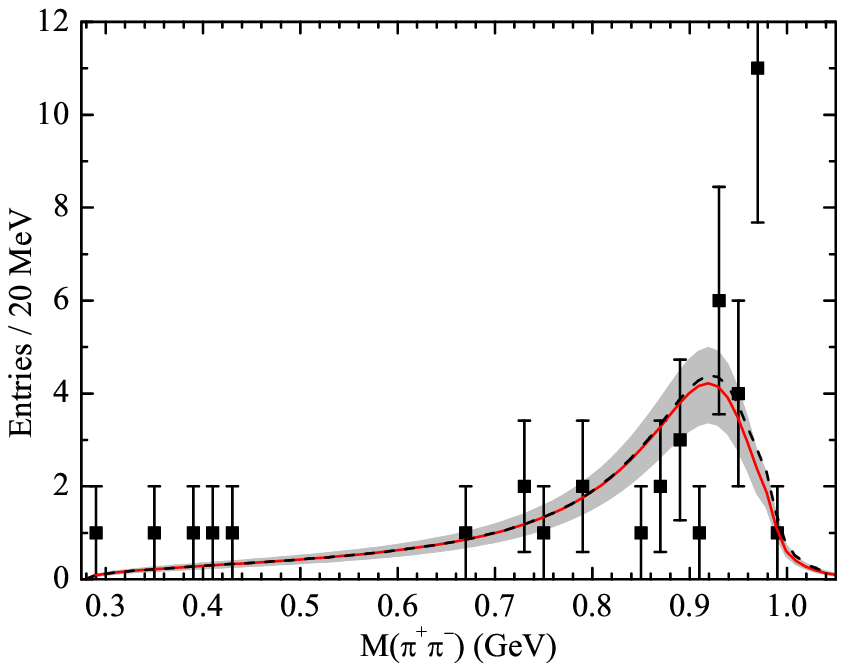}\hspace{2pc}
\includegraphics[width=0.52\textwidth]{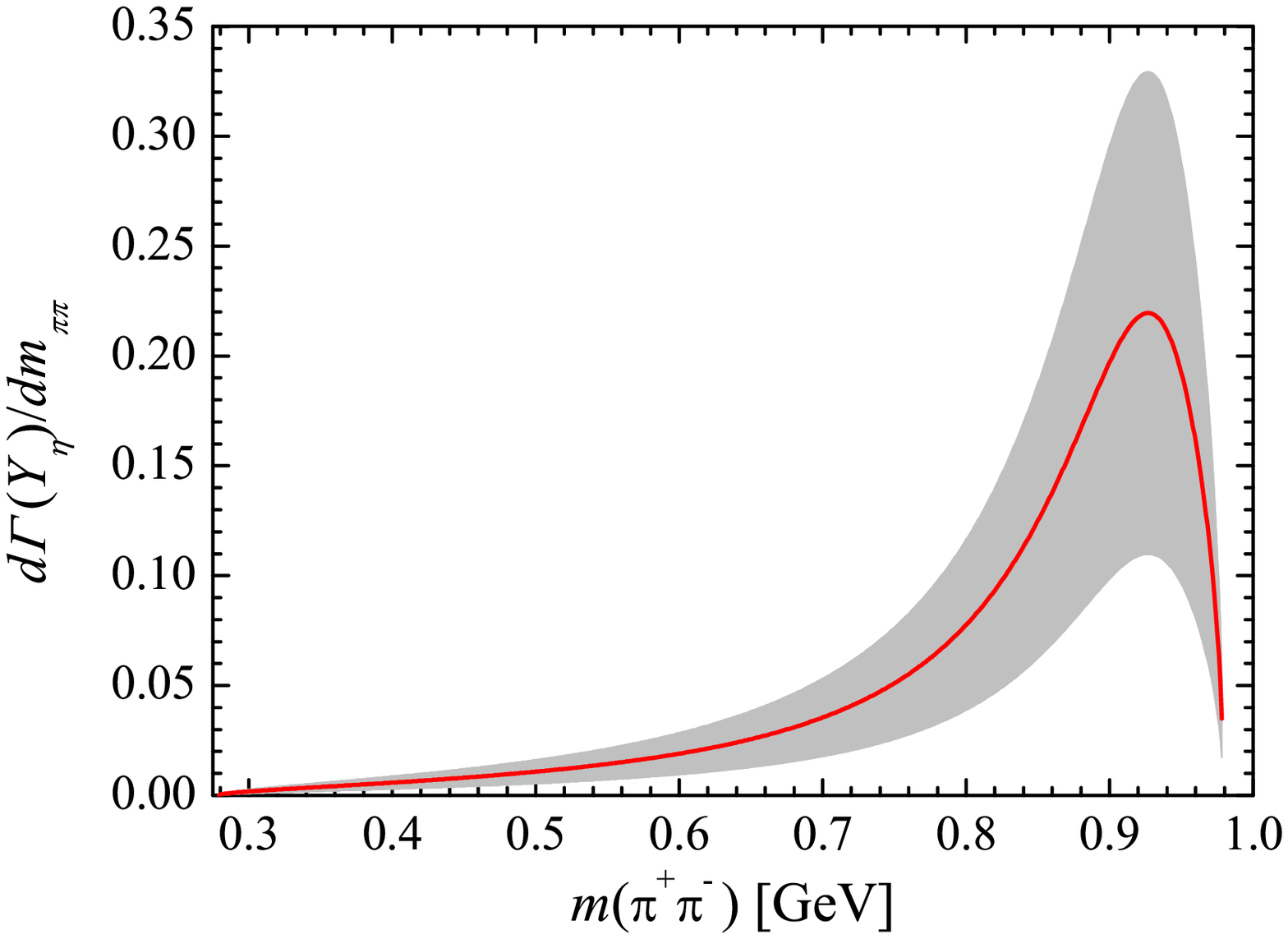}
\caption{\label{fig:Yeta}Left: Prediction for the $\pi^+\pi^-$ mass
  distribution in the molecular picture of the $Y(4660)$ (solid line: central
  parameters, shaded area: theoretical uncertainty) compared to the data 
  \cite{Yexp}. Right: Prediction for the $\pi^+\pi^-$ mass
  distribution in the molecular picture of the $Y_\eta$.}
\end{minipage}
\end{figure}

Now let me move to the heavy quark sector of $c$ and $b$ quarks, 
where heavy quark EFT is applicable. In the heavy quark limit $m_Q\to \infty$,
the QCD Lagrangian becomes independent of spin and flavor at leading order.
Due to this symmetry, there are spin multiplets of both
heavy mesons and heavy quarkonia, as e.g. the $\{D,D^*\}$ and
$\{\eta_c,J/\psi\}$. The masses of the members within the same spin
multiplet would be degenerate in the heavy quark limit. Recently,
this symmetry was extended to possible heavy meson molecules~\cite{Guo:2009id},
which were observed in recent years. Heavy meson molecules are bound
states consisting of a heavy meson/heavy quarkonium and a light
hadron, or two heavy mesons. It can be argued that the hyperfine 
splitting within a heavy quarkonium spin multiplet is untouched 
by their interactions with light mesons.  As a result, a bound state 
of a heavy quarkonium and light hadrons, would have
partner(s) whose components are the same light hadrons and the
spin-multiplet partner(s) of the same heavy quarkonium.
In particular, if the $Y(4660)$ is a $\psi'f_0(980)$ bound 
state~\cite{Guo:2008zg}, then there should exist an $\eta_c'f_0(980)$ 
bound state,  called $Y_\eta$, with its mass given as
\begin{equation}
\label{eq:my} M_{Y_\eta} = M_{Y(4660)} -
\left(M_{\psi'}-M_{\eta_c'}\right) = 4616^{+5}_{-6}~{\rm MeV}.
\end{equation} 
The quantum numbers of such a state are $J^P=0^-$.
The molecular picture of the  $Y(4660)$ describes naturally the
invariant $\pi^+\pi^-$ mass distribution measured in the reaction
$e^+e^-\to \gamma_{\rm ISR}\pi^+\pi^-\psi'$ as shown in the left
panel of Fig.~\ref{fig:Yeta}. For other recent work on the molecular
nature of the $Y(4660)$, see Refs.~\cite{Hagen:2010wh,Guo:2010tk}.
Using heavy quark spin symmetry,
one can then predict the analogous $\pi^+\pi^-$ mass spectrum of the
$Y_\eta$ decay (right panel of  Fig.~\ref{fig:Yeta}). Its eventual 
measurement would serve as an excellent testing ground of the molecular 
nature of both the $Y(4660)$ as well as its spin-partner, the $Y_\eta$.
Exciting times are ahead of us in view of the many new data from BESIII
at the BEPC, from the B-factories and, in the future, from PANDA at the HESR.
 
\section{Experimental spin physics at J\"ulich: a few selected high-lights}

Next, I would like to review some recent work on experimental spin
physics performed at J\"ulich at the Cooler Synchrotron COSY. Again, this is
a very subjective choice but is intended to show that COSY can be considered
as ``the spin machine''.
\begin{itemize}
\item[$\bullet$] COSY provides polarized proton and deuteron beams impinging
on unpolarized and polarized targets. This allows to perform a variety
of fundamental investigations. Besides the pion production experiments
already mentioned, charge symmetry breaking 
in $\vec d \ d\to \alpha  \pi^0$ (separation of s- and p-waves)
will be studied with WASA-at-COSY. Furthermore,
the reaction $\vec p \ p \to pK\Lambda$ will be investigated at TOF with the
aim of a determination of the singlet and triplet $\Lambda p$ scattering lengths.
\item[$\bullet$] The SPIN@COSY collaboration has performed many
  ``spin-gymnastics'' experiments. Of particular importance for a precision
 measurement of the $\eta$ mass from the threshold cross section of the
 reaction $pd \to$ $^3{\rm He} \ \eta$  was the recent
 determination of the COSY beam momentum with an unprecedented accuracy,
 $\Delta p/p < 10^{-4}$ at 3~GeV beam momentum, by depolarizing the beam 
 through the use of an artificially induced spin resonance \cite{Goslawski:2009vf}.  
\item[$\bullet$] The PAX collaboration  works on methods to polarize
 antiprotons. It was recently suggested that this can be done in spin-flip
 reactions \cite{Walcher:2007sj}, although the underlying theoretical 
 calculations were challenged early \cite{Milstein:2008tc}. 
 At COSY, the inverse reaction - namely the depolarization of
 protons through interactions with electrons - was measured and it was
 shown that the spin-flip mechanism advocated in  \cite{Walcher:2007sj} 
 is not a viable  tool to polarize antiprotons~\cite{Oellers:2009nm}.
\item[$\bullet$] In the context of the possible measurements of the 
 proton and the deuteron electric dipole moment in storage rings (see
 the next section), the EDM collaboration performs a variety of tests at
 COSY, such a setting up and testing polarimeters to be used in EDM
 experiments \cite{ES}.
\end{itemize}

\section{The nucleon electric dipole moment}
\begin{figure}[h]
\includegraphics[width=20pc]{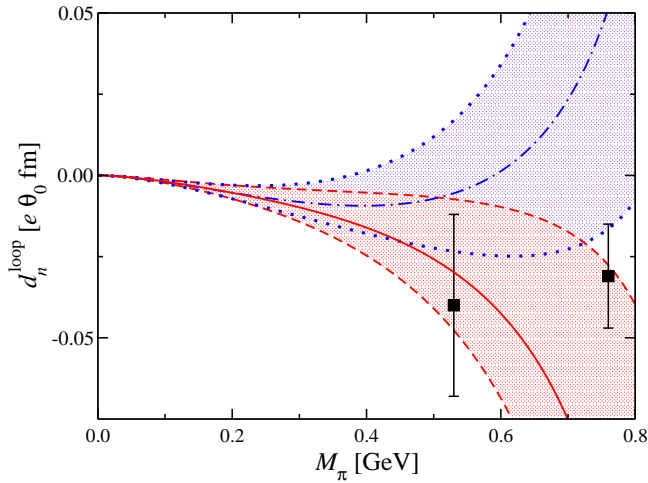}\hspace{2pc}%
\begin{minipage}[b]{14pc}\caption{\label{fig:dn} Chiral extrapolation of the
loop contribution to $d_n$. Dot-dashed line (with dotted borderlines): leading
one-loop result, solid line (with dashed borderlines): full one-loop result.
The data are from Ref.~\cite{Shintani:2008nt}. Note, however, that the lattice
data are at too high pion masses for using the chiral extrapolation formula
derived in~\cite{Ottnad:2009jw}.}
\end{minipage}
\end{figure}

The neutron electric dipole moment (nEDM) is a sensitive probe of CP
violation in the Standard Model and beyond. The current experimental
limit $d_n \leq 2.9 \cdot 10^{-26}\, e \, {\rm cm}$
is still orders of magnitude larger than
the Standard Model prediction due to weak interactions. However, in QCD 
the breaking of the $U(1)_A$ anomaly allows for strong CP violation,
which is parameterized through the vacuum angle $\theta_0$. Therefore,
an upper bound on $d_n$ allows to constrain the magnitude of  $\theta_0$. 
New and on-going experiments with ultracold neutrons
strive to improve these bounds even further, see e.g.~\cite{{Lamoreaux:2009zz}} for 
a very recent review. On the
theoretical side, first full lattice QCD calculations of the neutron and the proton
electric dipole moment  are becoming available~\cite{Berruto:2005hg,Shintani:2008nt,Aoki:2008gv}.
These require a careful study of the quark mass dependence of the nEDM
to connect to the physical light quark masses. In addition, there is a BNL
proposal to measure the proton and the deuteron EDM in a storage ring \cite{GO}
and there are also plans to build such type of machine in J\"ulich \cite{FR}.
It is thus of paramount interest to improve the existing
calculations of these fundamental quantities in the framework of 
chiral perturbation theory, as recently done in Ref.~\cite{Ottnad:2009jw}.
The calculation is based on $U(3)_L\times U(3)_R$ covariant baryon chiral
perturbation and includes {\sl all} diagrams contributing at one loop order. 
To be specific, let me first give some basic definitions.
The nucleon matrix element of the electromagnetic current in the presence of strong 
CP-violation is given by
\begin{equation}
\langle{p'} | J_{em}^{\nu} |{p}\rangle =
\bar u(p')\left[
\gamma^{\nu}F_1\left(q^2\right) - \frac{i}{2m}\sigma^{\mu\nu}q_\mu F_2\left(q^2\right) 
- \frac{1}{2m} \sigma^{\mu\nu}q_\mu\gamma_5 F_3\left(q^2\right)+\ldots\right] u(p)
\end{equation}
with $q_\mu = (p'-p)_\mu$. Here, $F_1$ and $F_{2}$ denote the P-, CP-conserving Dirac and Pauli 
form factors, $m$ is the mass of the nucleon, and $F_3$ the P- and
CP-violating electric dipole form factor. The ellipsis denotes the anapole
form factor that we do not consider here. The electric dipole moment of the
neutron/proton and the corresponding electric dipole radii follow as:
\begin{equation}
d_{n,p} = \frac{F_{3,n,p}(0) }{2m}~, ~~  \left\langle r_{ed}^2 \right\rangle 
= 6 \frac{d F_3(q^2)}{dq^2}\bigg|_{q^2=0} \ .
\end{equation}
I summarize briefly the pertinent results of the study presented 
in Ref.~\cite{Ottnad:2009jw}. First, a bound for
the vacuum angle could be given, $ \left| \theta_0 \right| \lesssim 2.5 \times
10^{-10}$. Second, the chiral expansion of the electric dipole radii
takes the form (where $\delta$ denotes a genuine small parameter)
\begin{eqnarray}
\langle r^2_{ed}\rangle_n = -20.4 \, \left[1 - 0.67 + \Order(\delta^2)\right]\,
\theta_0 \, e \, {\rm fm}^2~,\nonumber\\
\langle r^2_{ed}\rangle_p = +20.9 \, \left[1 - 0.70 + \Order(\delta^2)\right]\,
\theta_0 \, e \, {\rm fm}^2~,
\end{eqnarray}
where the large contribution of the NLO correction can be traced back to
an enhancement of the pion loop effect due to an extra factor of $\pi$,
similar to what was observed in the analysis of the isospin-violating
nucleon form factors~\cite{Kubis:2006cy}. Third, to
compare results from (two-flavor) lattice QCD at unphysical quark
masses with predictions from chiral perturbation theory, it is necessary to
perform an extrapolation of the analytic results in the pion mass 
(see \cite{Ottnad:2009jw} for the details). In Fig.~\ref{fig:dn} 
we show the resulting pion (quark) mass dependence
of the loop contribution to the neutron electric dipole moment in comparison to the 
available data points from two-flavor lattice QCD~\cite{Shintani:2008nt}.
It is interesting to see that the complete one-loop calculation (solid line)
reproduces the
trend of the lattice data (the order of magnitude and the global sign) 
even without the unknown tree contribution, quite in contrast to the
leading one-loop contributions (dot-dashed line).
However, only below pion masses of the order of about 500~MeV the  
corrections are sufficiently small for a  stable chiral extrapolation
as indicated by the theoretical uncertainties also shown in Fig.~\ref{fig:dn}.
For other recent work on the electric dipole form factor of the nucleon,
see~\cite{deVries:2010ah,Mereghetti:2010kp}.

It is also of interest to analyze the deuteron EDM, which has one- and
two-body contributions. As shown in Ref.~\cite{Liu:2004tq}, the two-body corrections
are generated from the induced P-wave in the deuteron wave function
(polarization effects) as well as from pion-exchange currents. In particular,
one finds a very different sensitivity to quark or chromomagnetic EDMs
compared to the nucleon. For details, I refer to~\cite{Liu:2004tq}. It 
certainly would be worthwhile to repeat this calculation in the framework of 
chiral nuclear EFT.

\section{Ab initio calculation of hadron resonances}

Arguably the most outstanding problem in QCD is to understand its spectrum.
Questions in this context are: why does QCD mostly produce hadrons in
terms of $\bar q q$ and $qqq$ states? What is the nature of the puzzling
$X,Y,Z$ states and the related charm-strange mesons? Where are the
exotica and the glueballs? Experimental facilities world-wide produce
a cornucopia of high-precision data relevant to spectroscopy, often involving
polarized beams and/or targets. In theory, lattice QCD is the premier tool 
to investigate the (low-lying) spectrum. In the light quark ($u,d,s$) sector,
there has been tremendous progress with simulations at almost physical quark
masses, see e.g. \cite{Durr:2008zz,Aoki:2008sm,Baron:2010bv,Engel:2010my}.
However, most hadrons are resonances - and therefore it is of utmost
importance to be able to calculate their decay properties (partial and
total width(s)). 

I will concentrate here on the most significant baryon
resonance, the $\Delta(1232)$. It is a well isolated resonance with its
pole in the complex energy plane located at $(E,i\Gamma) = (1210, i50)\,$MeV. 
The $\Delta$ was first observed in pion-nucleon scattering and is known to
dominate the photonuclear response. It is also amenable to a systematic EFT
treatment, provided one counts the delta-nucleon mass splitting as a small
parameter. The question how to calculate the width of resonance in a finite
volume was in principle answered by L\"uscher (and others) - in the vicinity
of a narrow, well-separated resonance the two-particle energy levels exhibit
a marked volume dependence in form of an avoided level crossing. This method
was applied to synthetic finite volume data for pion-nucleon scattering in the
$P_{33}$ partial wave in Ref.~\cite{Bernard:2007cm}. In fact, the avoided
level crossing is washed out, but still a precise measurement of the volume
dependence of the difference of the first two energy levels would allow
to extract the delta-pion-nucleon coupling constant $g$ and thus the width of the
$\Delta$. A different way of representing the same physics was discussed in
Ref.~\cite{Bernard:2008ax} which in principle works for both narrow as well
as broad resonances. This was demonstrated very nicely in a toy model by
Morningstar~\cite{Morningstar:2008mc} (see also Ref.~\cite{Giudice:2010ch}). 
Furthermore, the finite volume
corrections for a decaying $\Delta$ have been worked out in \cite{Bernard:2009mw}
and successfully applied in combined fit to the data of the ETM collaboration
for the pion mass dependence of the nucleon and the delta masses \cite{Alexandrou:2008tn}. 

While these methods are working in principle, not sufficiently many 
lattice data for large enough volumes exist to cleanly extract the 
$\Delta$ width based on the energy level difference or the
statistical method. However, for elastic two-body resonances, 
like  the $\Delta$, a simpler method, originally proposed by L\"uscher~\cite{Luscher} and
Wiese~\cite{Wiese}, is based on computing the scattering phase shift
in the infinite volume from the volume dependence of the energy levels
of the lattice Hamiltonian. The delta resonances is practically a 
two-body state, and phenomenologically its scattering phase shift 
$\delta$ is very well described by the effective range formula
\begin{equation}
\frac{k^3}{W}\,\cot\,\delta(k) = \frac{6\pi}{g^2}\,
\left(m^2-W^2\right)\;,
\label{effective}
\end{equation}
where $k=|\vec{k}|$ is the constituents center-of-mass momentum,
$W=\sqrt{k^2+m_1^2}+\sqrt{k^2+m_2}$ and $g$ is the coupling
constant, which is related to the width of the resonance by
\begin{equation}
\Gamma=\frac{g^2}{6\pi} \frac{k^3}{m^2}\, .
\end{equation}
The phase shift $\delta$ passes through $\pi/2$ at the physical mass,
i.e. $W^2=m^2$.
In the case of noninteracting particles the possible energy levels in
a periodic box of length $L$ are given by
\begin{equation}
W=\sqrt{k^2+m_1^2}+\sqrt{k^2+m_2^2}\;,
\label{energy}
\end{equation}
where $k=2\pi |\vec{n}|/L$, $\vec{n}$ being a vector with components $n_i \in
\mathbb{N}$. In the interacting case, the energy levels are still given by
(\ref{energy}), but now $k$ is the solution of~\cite{Luscher,Bernard:2008ax}
\begin{equation}
\delta (k) = \arctan\left\{\frac{\pi^{3/2}
    q}{\mathcal{Z}_{00}(1,q^2)}\right\}\: \mbox{mod}\; \pi\;, \quad
    q=\frac{kL}{2\pi}\;, 
\label{phase}
\end{equation}
where $\mathcal{Z}_{00}$ is a generalized zeta function,
\begin{equation}
{\cal Z}_{00} (1;q^2) = \frac{1}{\sqrt{4\pi}} \,\sum\limits_{\vec{n} \in \mathbb{Z}^3}
\displaystyle\frac{1}{{\vec n\,}^2 - q^2}~.
\end{equation}
That is to say, each energy value $W$, computed on the periodic
lattice at some fixed values of $m_1, m_2$, gives rise to a certain
momentum $k$.
The scattering phase at this momentum and pion mass is given by
Eq.~(\ref{phase}). Fitting $\delta$ to the effective range formula
Eq.~(\ref{effective}) then allows us to estimate the mass and width of the
actual resonance. 
Eq.~(\ref{phase}) holds for vanishing total momenta. For
nonvanishing momenta different zeta functions or combinations of zeta
functions apply~\cite{Gottlieb}.

\begin{figure}[!ht]
\includegraphics[width=20pc]{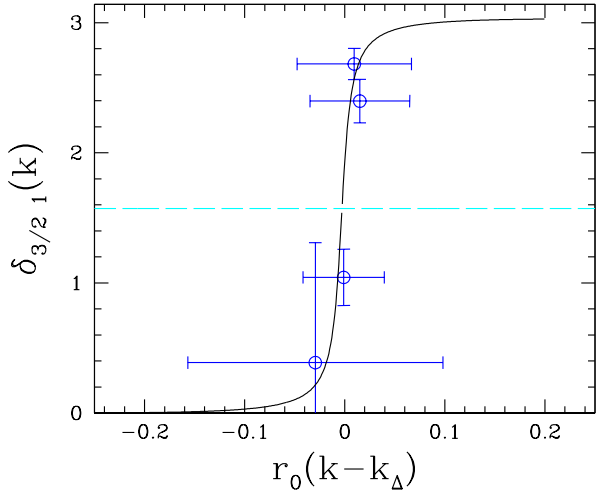}\hspace{2pc}%
\begin{minipage}[b]{14pc}\caption{\label{fig:del}
The $\pi N$ phase shift of the $\Delta$ channel as a function $k$.
The solid line is the phase generated from the physical mass
and width of the $\Delta$-resonance. The dashed horizontal line
indicates the $90^\circ$ crossing of the phase.}
\end{minipage}
\end{figure}

The $\pi N$ phase shift of the $\Delta$ channel is more difficult to
compute than the corresponding $\pi\pi$ $I=J=1$ phase that contains the
$\rho$ (see e.g. Refs.~\cite{Aoki:2007rd,Gockeler:2008kc,Feng:2010es,Frison:2010ws})
because the phase space is much smaller at comparable pion
masses. Nonetheless, together with the QCDSF collaboration, we are presently 
investigating the $I=3/2$ P-wave in $\pi N$ scattering, using a 
$N_f=2$ clover action that consists of the plaquette gluon action 
together with nonperturbatively $O(a)$ improved Wilson (clover) fermions.
In Fig.~\ref{fig:del} we show some first results, combining calculations
on $32^3\times 64$ and $40^3\times 64$ lattices at $\beta=5.29,
\kappa=0.13632, a = 0.075\,$fm and on $40^3\times 64$ and $48^3\times 64$ 
lattices at $\beta=5.29, \kappa=0.1364, a = 0.06\,$fm. So far the errors 
are relatively large,  which calls for higher statistics. To trace the 
phase shift over a  sufficiently large range of values
around $\delta=\pi/2$ with high precision, calculations need to be
formed on larger lattices and with nonvanishing total momenta. Such 
simulations are presently being performed. Still, as indicated by the solid line
that is generated with the physical mass and width of the $\Delta$,
we definitely are on the right track~\cite{MRS}.

\section{Summary \& outlook}

Spin physics is and will be an exciting field of research. Much progress
has been made through developments in accelerator, experimental and
theoretical physics. Here, I have focused on some recent developments
in the sector of  non-perturbative QCD, stressing the fruitful interplay
between theory and experiment. On the theoretical side, much progress 
has been made in the use and application of effective field theories and/or
lattice simulations. While lattice simulations are quickly approaching
the limit of physical pion masses, chiral extrapolations and finite
volume EFTs are nevertheless required to fully access the underlying physics,
as best exemplified for the case of instable particles on the lattice.
Similarly, the rich spectrum in the charm sector measured at colliders
and B-factories requires a thorough investigation of possible  mechanisms
to generate hadrons, such as the formation of hadronic molecules 
discussed in Sec.~2.
Finally, precision measurements combined with accurate theoretical tools
allow for many fine test of the chiral QCD dynamics, as discussed here
for the determination of the low-energy constant $D$, that appears in
a cornucopia of low-energy processes. The exciting possibility of measuring
the electric dipole moment of the proton and light nuclei should lead to
refined theoretical investigations, especially using chiral nuclear
effective field theory. Experiments involving polarization will continue
to deliver accurate data that will  challenge our understanding 
of the Standard Model and might give possible signatures of physics beyond it. In this
context, it would be very important to understand the mechanism beyond the
polarization of antiprotons, which, if implemented within the HESR complex
at FAIR, would open exciting new possibilities to explore the physics of
charm and light quarks in hadronic and nuclear systems.

\section*{Acknowledgements}
I thank the organizers for giving me this opportunity and all my
collaborators, who have contributed to my understanding of the
issues discussed here. Partial financial support by the Helmholtz 
Association through funds provided to the virtual institute 
``Spin and strong QCD'' (VH-VI-231), by the European Community-Research 
Infrastructure Integrating Activity ``Study of Strongly Interacting Matter''
(acronym HadronPhysics2, Grant Agreement n.~227431) under the Seventh
Framework Programme of the EU,  by DFG (SFB/TR 16, ``Subnuclear 
Structure of Matter'') and by  BMBF (grant 06BN9006) is gratefully acknowledged.

\pagebreak

\end{document}